\documentclass[conference,peer-reviewed]{aesconf}

\graphicspath{{./}{figures/}}

\usepackage[utf8]{inputenc}

\usepackage{microtype}

\usepackage[numbers,square]{natbib}

\usepackage{booktabs}
\usepackage{color}
\usepackage{url}
\usepackage{graphicx}
\usepackage{array}
\usepackage{multirow}
\usepackage{textcomp}
\usepackage{float}
\usepackage{amsmath}

\newcommand{\SSS}{\texttt{SS}}
\newcommand{\MS}{\texttt{MS}}

\title{Room Impulse Response Estimation in a Multiple Source Environment}

\author[]{Kyungyun Lee}
\author[]{Jeonghun Seo}
\author[]{Keunwoo Choi} 
\author[]{Sangmoon Lee}
\author[]{Ben Sangbae Chon}
\affil[]{Gaudio Lab, Inc., Seoul, South Korea} 

\correspondence{Ben Sangbae Chon}{bc@gaudiolab.com}

\lastnames{Lee, Seo, Choi, Lee, Chon}

\shorttitle{RIR Estimation in a Multiple Source Environment}

\savebox{\AEStop}{%
	\begin{minipage}{\textwidth}%
		\rule{\textwidth}{1.5pt}\\%
		\\%
		\begin{minipage}[c][\iftoggle{convention}{3.2cm}{3.7cm}][t]{0\textwidth}%
			\includegraphics[width=20mm]{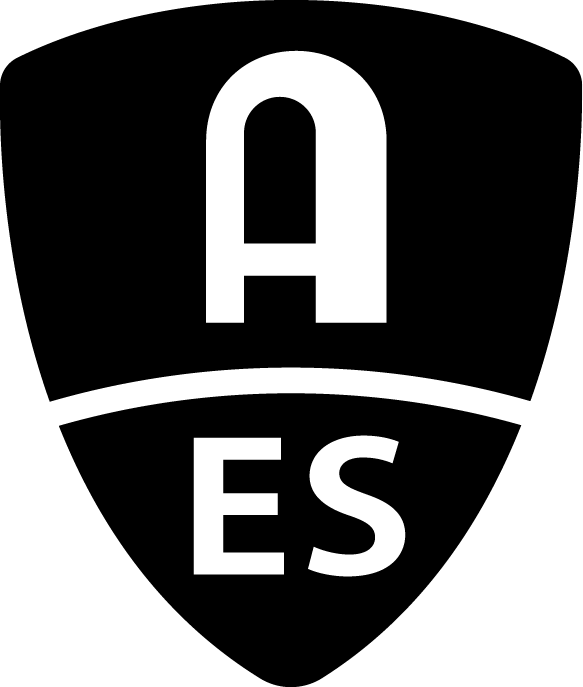}%
		\end{minipage}%
		\begin{minipage}{\textwidth}%
			\sffamily%
			\begin{center}%
				\LARGE Audio Engineering Society\\%
				\iftoggle{express_paper}{%
				\hspace{3mm}\fontsize{36}{38pt}\selectfont Convention Express\\Paper \AESExpressPaperNumber\\%
				}{%
				\iftoggle{convention}{%
				\fontsize{36}{38pt}\selectfont Convention Paper\\%
				}{%
				\fontsize{36}{38pt}\selectfont Conference Paper\\%
				}}%
				\vspace{0.2cm}%
				\large Presented at the \AESConferenceNumber \iftoggle{convention}{Convention\\}{\AESConferencePrefix Conference on\\}%
				\iftoggle{convention}{}{\AESConferenceTopic\\}%
				\AESConferenceDate\ifx\AESConferenceLocation\empty\else, \AESConferenceLocation\fi%
			\end{center}%
		\end{minipage}\\%
		\vspace{0.2cm}\\%
		\begin{minipage}{\textwidth}%
			\rmfamily\itshape\small	\AESLegalTextPrefix\ \AESLegalText%
		\end{minipage}\\%
		\\%
		\rule{\textwidth}{1.5pt}%
	\end{minipage}%
}

\begin{document}

\twocolumn[
\maketitle 

\begin{onecolabstract}
In real-world acoustic scenarios, there often are multiple sound sources present in a room. 
These sources are situated in various locations and produce sounds that reach the listener from multiple directions. The presence of multiple sources in a room creates new challenges in estimating the room impulse response (RIR) as each source has a unique RIR, dependent on its location and orientation. 
Therefore, issues of determining which RIR should be predicted and how to predict it arise, when the input signal is a mixture of multiple reverberated sources. 
To address these, we propose a new task of predicting a "representative" RIR for a room in a multiple source environment and present a training method to achieve this goal. In contrast to the model trained in a single source environment, our method shows robust performance, regardless of the number of sources in the environment. 
\end{onecolabstract}
]

\section{Introduction}


The acoustic properties of a room are described by its room impulse response (RIR), which depends on factors such as room geometry, materials, and object layout. Having RIR in hand is critical for various applications, including virtual and augmented reality \cite{jot2016augmented, jot2021rendering, gariroom, rafaely2022spatial}, dereverberation \cite{lebart2001new}, and speech recognition \cite{ko2017study}. However, obtaining RIRs through physical measurements is an arduous and time-consuming task, motivating research on methods for estimating RIR. 
One area of research on RIR estimation is blind estimation, which directly predicts RIR from reverberated source recordings without any additional information of the room \cite{gamper2018blind, steinmetz2021filtered, ratnarajah2022towards}. However, most previous methods only consider a single sound source in a room, while in reality, rooms often have multiple sound sources at various locations.

\begin{figure}[h]
\begin{center}
\includegraphics[width=\columnwidth]{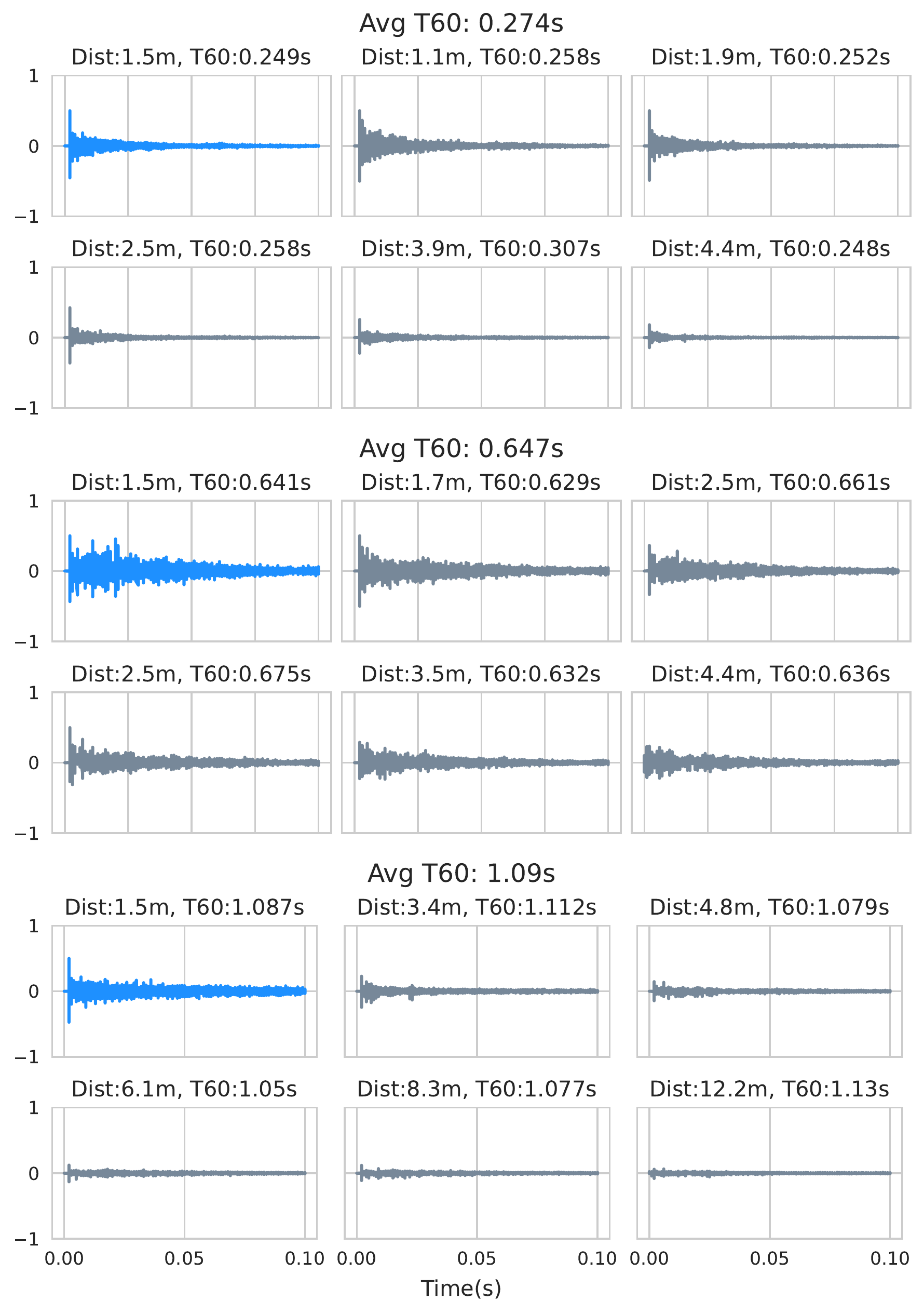}
\caption{ Although the T60 values for RIRs measured in the same room are similar, the waveform shapes vary depending on the source-receiver distance and direction. The representative RIR is in blue.
} 
\label{fig:rirs_per_room}
\end{center}
\end{figure}

As shown in Figure \ref{fig:rirs_per_room}, RIRs vary based on the source-receiver position and orientation, posing a challenge in determining which RIR or portion of the RIR should be estimated in a multiple source environment. Ultimately, the solution will depend on the specific task at hand, which may require estimation of RIRs of all sources or a single RIR that captures general information about the room.

In this work, we introduce a task of estimating a single, central RIR in a multiple source environment. Our key assumption is that a single RIR can serve as an effective representation of the acoustic characteristics of a room. This hypothesis is grounded in the fact that humans have the ability to distinguish whether different sound sources come from the same room, despite variations in source locations \cite{helmholz2022towards}. Additionally, it is well known that while the direct and early reflections of RIR are influenced by the source-receiver position, the late reverberation provides a more comprehensive understanding of the room's acoustics \cite{griesinger1997psychoacoustics, kuttruff2016room}. Based on these observations, there exists some shared acoustical features among RIRs that originate from the same room.
In the following sections, we introduce the training method for estimating representative RIRs, including data processing procedures, and present evaluation metrics for multiple source environments. We also share how we post-processed the estimated monaural RIRs to obtain binaural filters for spatial audio rendering use cases. 

\subsection{Related works}
Research on blind estimation of RIR or room acoustic parameters from reverberated sources in single source environments has been extensive, employing methods from signal processing \cite{antsalo2001estimation, crammer2006room, crocco2015room}  to deep learning \cite{gamper2018blind, steinmetz2021filtered, ratnarajah2022towards, bryan2020impulse, yu2020room}. Previously, room acoustic parameters, such as T60 and direct-to-reverb ratio (DRR), were more commonly estimated, but recent advances in neural networks allowed direct RIR estimation in the time domain \cite{steinmetz2021filtered, ratnarajah2022towards}. Despite efforts to address the lack of data for training neural networks \cite{szoke2019building, bryan2020impulse}, data remains as a challenge, resulting in an active usage of synthetically generated data. 
There is even a smaller number of dataset for training in a multiple source environment \cite{peters2012name, carlo2021dechorate, tang2022gwa}, which requires measurements of RIRs at multiple locations in a room and in several thousand unique rooms. 

Another line of research is a multi-modal approach on estimating RIR from images, videos, or 3D meshes to render sound in a specific room \cite{tang2020scene, garg2021geometry, chen2022visual}. In addition, an extensive study involves synthesizing a RIR given source-receiver locations and room descriptions, such as room geometry and T60, which is especially useful for virtual reality applications \cite{ratnarajah2020ir, ratnarajah2022fast, richard2022deep, ratnarajah2022mesh2ir}.

\begin{figure}
\begin{center}
\includegraphics[width=.8\columnwidth]{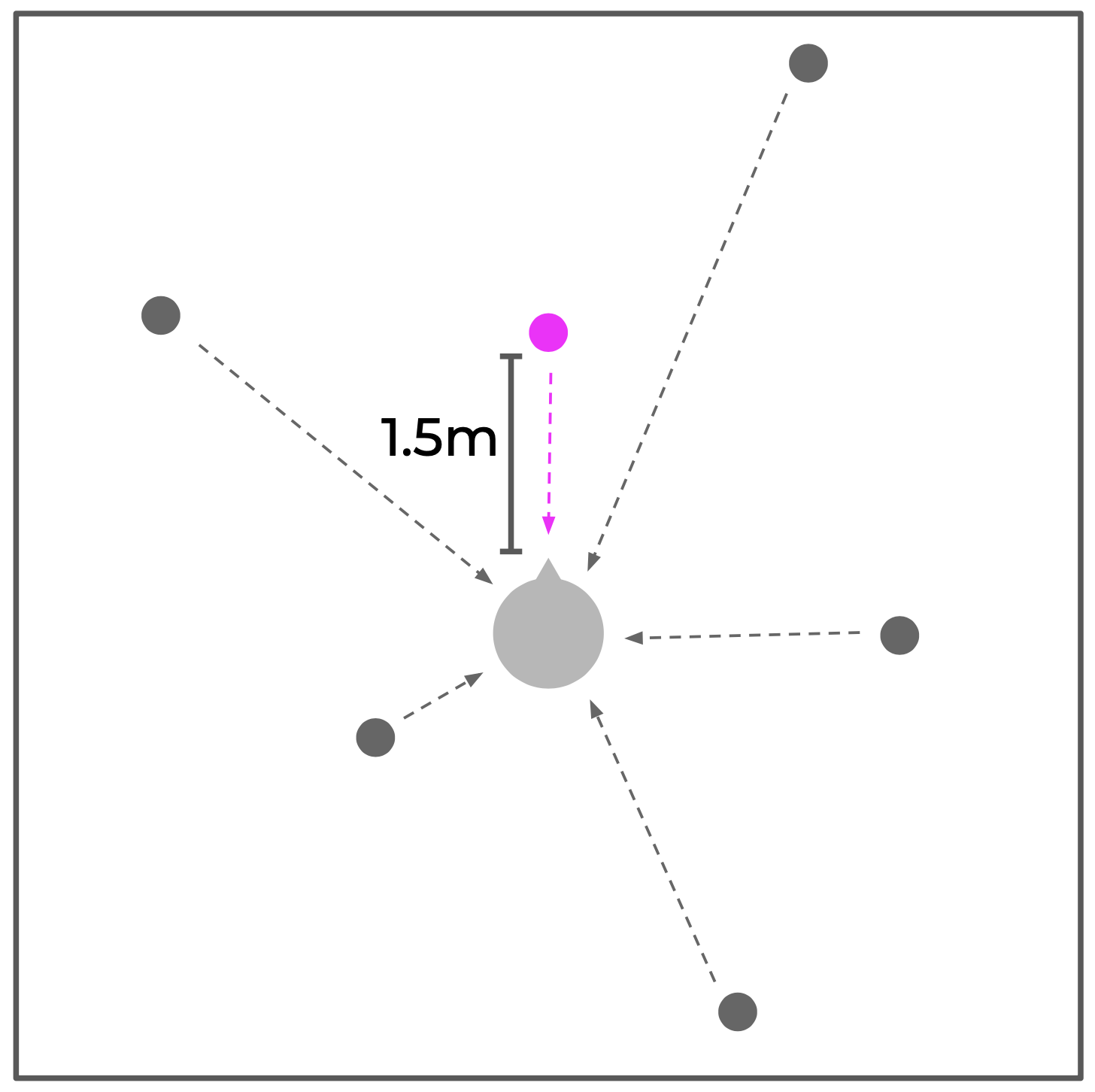}
  \caption{In a multiple source environment, there are many sources in the room (gray circles). The representative RIR is designated as the RIR from a source that is 1.5 meters away from the listener at the center (pink circle).}
  \label{fig:multiple_source_setup}
\end{center}
\end{figure}

\section{Methods}
In a multiple source environment, sources are located in various locations within the room and emitted sounds reach the listener from all directions (Figure \ref{fig:multiple_source_setup} gray circles). Regardless of how many sources there are and where each source is located, we define the representative RIR as the one that is located 1.5 meters away from the listener at center. We chose 1.5 meters as it approximates the typical conversational distance between people in a room. We imagine virtual conversations as a common use cases for AR/VR scenario. Furthermore, for spatial audio music production, it is recommended to render with a center speaker placed at 1.5 meters from the listener\footnote{https://professionalsupport.dolby.com/s/article/Dolby-Atmos-Music-Studio-Best-Practices?language=en\_US}. Therefore, we setup the system to always predict this central representative RIR, when given any multiple source reverberated recordings.

\begin{figure}[h]
  \includegraphics[width=\columnwidth]{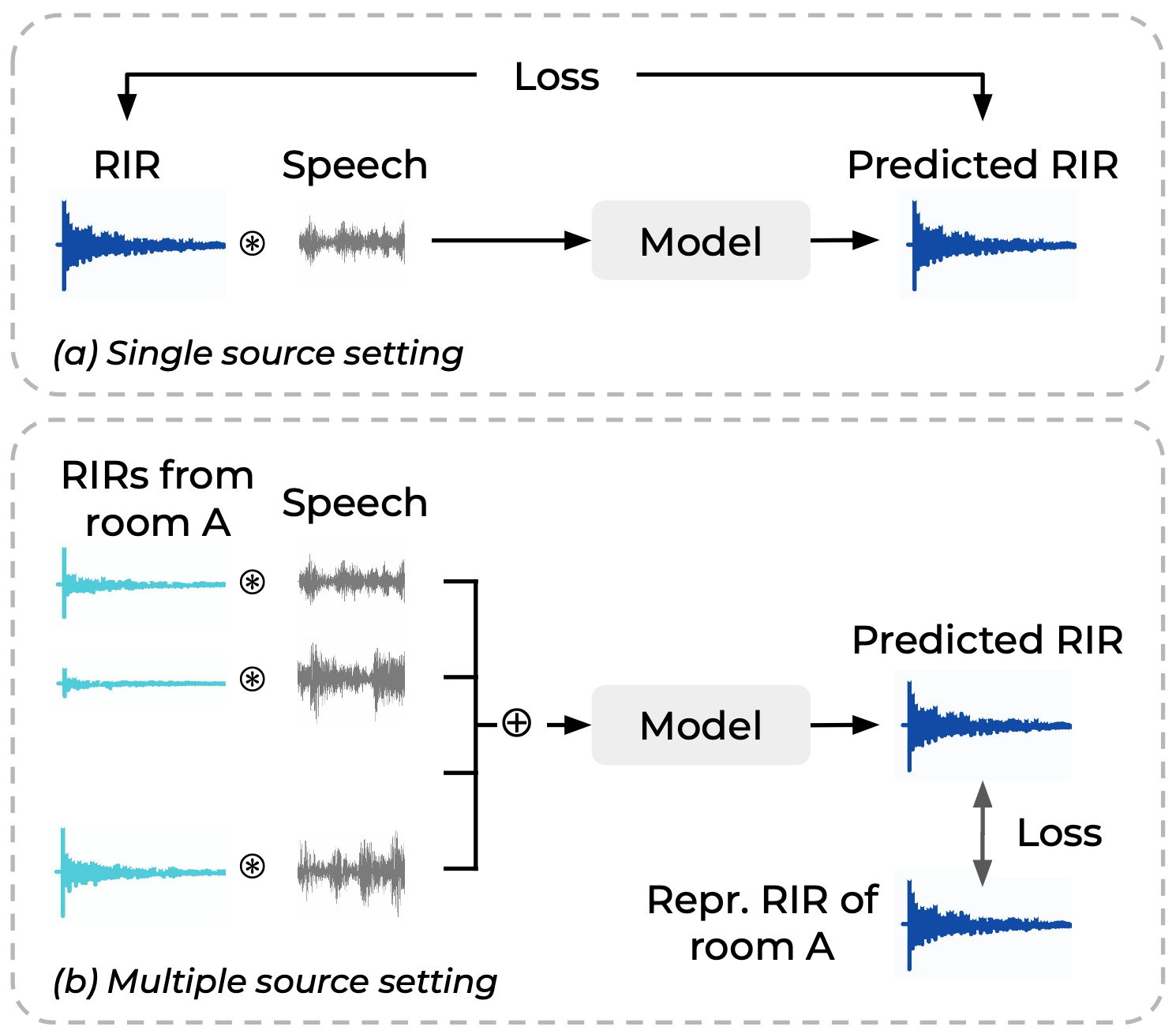}
  \caption{Single source (\SSS{}) vs. multiple source (\MS{}) training setup. In the \MS{} setup, the model is trained to predict the representative RIR, which may not be present in the RIRs utilized to produce the reverberated audio recording.}
  \label{fig:training}
\end{figure}



\subsection{Dataset} 

In order to train a model for multiple source scenarios, it is essential to have a dataset that comprises of RIRs measured at different locations within the room and from sufficiently large number of unique rooms. As listed in Table \ref{table:dataset}, we were able to obtain only 72 rooms with real measured RIRs. To compensate for the limited number of data points, we utilized the large synthetic RIR dataset, GWA \cite{tang2022gwa}, which contains over 6000 rooms and 10 RIRs per room, and also generated our own synthetic dataset, referred to as GRIR. 

GRIR generation process uses the room simulation method from the pyroomacoustics library \cite{scheibler2018pyroomacoustics} to generate multiple RIRs with random source positions in a predefined room. Generated room size ranges from $24.5 m^2$ to $4000 m^2$ and T60 value ranges from 0.1s to 1.5s. After generating a RIR with the library, we applied the following fine-tuning steps to render a more realistic RIR. First, we replaced the late reverberation part of the signal with a frequency-dependent exponentially decaying white noise \cite{moorer1979reverberation}. This is because the late reverberation generated using the library is sparse and unnatural. Then, we applied a random frequency-dependent decay pattern for all RIRs in each room to mimic the unique patterns of a room arising from the type of wall materials and objects in the room. With the addition of 11,000 rooms from GRIR, we have collected data of more than 12,000 unique rooms.

Example distribution of the dataset can be seen in Table \ref{table:test_dataset}. The data demonstrates that, although there are variations in RIRs within a room, similarities can also be observed. The T30 values, which are not heavily influenced by the source-receiver location, exhibit minor variations while the C50 values, which are more dependent on early reflections, display larger variations.
Anechoic speech signals used as source signals are obtained from DAPS \cite{mysore2014can} and VCTK \cite{yamagishi2019cstr} dataset.

\begin{table*}[h]
\centering
\begin{tabular}{*7c}
Dataset/room &  \multicolumn{2}{c}{T30 [sec]} & \multicolumn{2}{c}{EDT [sec]} & \multicolumn{2}{c}{C50 [dB]}\\ 
{} & $\mu$ & $\sigma$ & $\mu$ & $\sigma$ & $\mu$ & $\sigma$ \\ \hline \hline
BUT/Hotel\_SkalskyDvur\_Room112 & 0.448 & 0.12 & 0.406 & 0.119 & 6.332 & 3.235 
 \\ 
BUT/VUT\_FIT\_L227 & 1.942 & 0.082 & 2.618 & 0.52 & -7.543 & 4.41
 \\ 
BUT/VUT\_FIT\_Q301 & 0.613 & 0.219 & 0.489 & 0.216 & 5.475 & 5.457 
 \\ \hline
GWA/room0 & 0.804 & 0.058 & 0.704 & 0.142 & 1.863 & 2.743
 \\ 
GWA/room1 & 0.503 & 0.067 & 0.341 & 0.033 & 8.699 & 0.497
 \\ 
GWA/room2 & 0.428 & 0.048 & 0.415 & 0.07 & 6.542 & 1.179 
 \\ \hline
 GRIR/room0  & 0.274 & 0.022 & 0.235 & 0.038 & 11.07 & 1.21
 \\ 
 GRIR/room1
 & 0.647 & 0.018 & 0.859 & 0.109 & 1.466 & 1.733
 \\ 
GRIR/room2  & 1.09 & 0.028 & 1.214 & 0.061 & 0.237 & 0.709
 \\ 
\end{tabular}
\caption{RIR distributions of selected rooms in test data. T30 (reverberation time for sound to decay by 30dB), EDT (early decay time) and C50 (clarity of speech) are the average of each values computed at octave bands with center frequency 500 and 1000 Hz.
} 

\label{table:test_dataset}
\end{table*}

\subsection{Data pre-processing} 
Prior to training the model, we perform a pre-processing step to normalize the RIRs across datasets.
For each room in the dataset, we first identify the representative RIR. For datasets with source-receiver distance and/or orientation metadata, RIRs at a distance around 1.5m (often between 1.4m and 2.0m) and at center are nominated. However, many datasets do not provide distance and orientation information. In such cases, we select the RIR with the largest peak value, because it implies that this RIR has the smallest source-receiver distance and most datasets measure the closest RIR at distances around 1.0m. 
After the identification of the representative RIR, we compute a normalization factor, referred to as a \textit{room normalization factor}, which normalizes the representative RIR's absolute peak amplitude to 0.5. 
This normalization factor is then applied equally to the rest of the RIRs in the same room. As a result, each RIR in the room can be normalized, while conserving the DRR. Figure \ref{fig:rirs_per_room} shows various RIRs in 3 selected rooms after normalization applied. 



\begin{table*}[t]
\centering
\begin{tabular}{c|c|c|c|c|c}
Dataset name & Real RIR & \# of rooms & \# of RIRs per room (min $\sim$ max) & Total \# of RIRs & Train/Valid/Test \\ \hline 
ACE \cite{eaton2016estimation} & True& 7 & 2 & 14 & Test \\
ASH IR & True & 39 & 5 $\sim$ 72 & 729 & Train \\
BUT \cite{gamper2018blind} & True & 9 & 62 $\sim$ 465 & 2325 & Train, Test \\ 
C4DM & True& 3 & 130 $\sim$ 169 & 468 & Train\\ 
dEchorate\cite{carlo2021dechorate} & True & 8 & 30 & 1680 & Train\\
GRIR & False & 9831 & 4 $\sim$ 14 & 58922 & Train, Valid, Test\\ 
GWA \cite{tang2022gwa} & False & 1556 & 3 $\sim$ 10 & 6653 & Train, Valid, Test\\
IoSR & True& 4 & 37 & 148 & Train \\ 
OpenAir & True& 21 & 2 $\sim$ 41 & 182 & Train\\ 
REVERB & True& 2 & 12 $\sim$ 24 & 36 & Train \\ 
RWCP & True& 7 & 7 19 & 81 & Train \\ 
R3VIVAL & True& 8 & 34 & 272 & Train\\ \hline 
Train & - & 9168 & 2 $\sim$ 465 & 64452 & Train \\
Valid & - & 1138 & 5 $\sim$ 10 & 7677 & Valid \\
Test & - & 1254 & 2 $\sim$ 434 & 8490 & Test \\ 
\end{tabular}
\caption{Dataset used for training and testing} 
\label{table:dataset}
\end{table*}



\subsection{Model and Training Method}
To apply our proposed training method, we implemented the Filtered Noise Shaping model  \cite{steinmetz2021filtered}, which directly generates RIR in the time-domain. Given a reverberated signal of 2.74s, the model predicts a 1 second RIR. The major characteristic of the model is that it generates the early reflections and late reverberation separately. Specifically, the model constructs the late reverberation as a sum of 10 decaying filtered noise, while the early components are generated sample by sample. The boundary between two parts is predefined to be at 0.05s.

As depicted in Figure \ref{fig:training}, previous works have utilized a single source training setup (\SSS), in which the input reverberated signal contains only one source and the model had to predict the corresponding RIR. In our proposed multiple source training setup (\MS), the model is trained to estimate the representative RIR when given a mixture of multiple reverberated source signals. Sources are located at different positions and thus, have different RIRs. 
During training, we dynamically synthesize input signals as a combination of 1 to 6 reverberated sources. Each of the reverberated sources is a convolution of a randomly selected RIR from the given room with a random anechoic speech signal. 

The input and output format of the \MS{} method is same as that of the \SSS{}: both take reverberated source signal as an input and the predicted RIR as an output. The model is trained with the multi-resolution STFT loss \cite{yamamoto2020parallel}, which consists of the spectral convergence loss (1) and the spectral log-magnitude loss (2), where $||\cdot||_F$ is the Frobenius norm, $||\cdot||_1$ is the L1 norm,
and N is the number of STFT frames. As the name suggests, the multiresolution STFT loss is the sum of aforementioned losses over $R$ different STFT resolutions (3). We experimented with additional loss functions, such as GAN loss \cite{ratnarajah2022towards} and energy decay relief (EDR) loss \cite{ratnarajah2022mesh2ir}, but they did not provide perceptually significant improvement. 

\begin{align} \label{eq1}
L_{sc}(h, \hat{h}) &= \frac{||STFT(h) - STFT(\hat{h})||_F}{||STFT(h)||_F}  \\
L_{mag}(h, \hat{h}) &= \frac{1}{N}||\log(STFT(h)) - \log(STFT(\hat{h}))||_1  \\
L_{stft} &= \sum^{R} L_{sc_r}(h, \hat{h}) + L_{mag_r}(h, \hat{h})
\end{align}

As in \cite{steinmetz2021filtered}, we use the sampling rate of 48kHz. Both the \SSS{} and \MS{} methods are trained on 2 GPUs each with a batch size of 32. Training was stopped after 600 epochs. When training the \MS{}, we used a checkpoint from the \SSS{} at 100 epochs and continued the training in the \MS{} setup. The initial learning rate was set to 5e-5. 

\subsection{Post-processing for binaural rendering} 
To test the practicality of the trained model in spatial audio applicationss, we extended the predicted monoaural RIRs need to BRIRs.
To this end, we posited that updating only the late reverberation with our predicted RIR, while retaining the direct and early reflection components from a HRIR would sufficiently represent various acoustical environments.
Our internal listening test indeed demonstrated that modifying the late reverberation alone was adequate to discern between different rooms.
Thus, we binauralize the late reverberation following the binauralization method from \cite{porschmann2017binauralization}, which convolves the segments of late reverberation with a set of binaural noise and sums with overlap-add. 
Then, we simply combine the synthesized late component with the publicly available CIPIC HRIR filters \cite{algazi2001cipic} as our internal application only requires BRIRs for a fixed location of sources. 
Since the predicted RIR has a maximum peak amplitude of 0.5 and the HRIR has that of 0.6, we rescale the predicted RIR to match HRIR. Two filters are combined using the Hann window. 
Note that the evaluation of the model is only done on monaural RIRs. 


\section{Results and Discussion}

\subsection {Objective evaluation}

\begin{table*}[h]
\centering
\begin{tabular}{ l|c|c|c|c|c|c } 

Test dataset & \# of Rooms & Real RIR &  Model & STFT Loss & T60 Loss & DRR Loss \\
\hline
\multirow{2}{6em}{ACE} & \multirow{2}{6em}{9} &  \multirow{2}{6em}{Yes} & \SSS{} & 2.49 & 0.092 & 224.17 \\ 
& & &  \MS{} & \textbf{2.072} & \textbf{0.081} & \textbf{169.12} \\ \hline
\multirow{2}{6em}{BUT} &  \multirow{2}{6em}{3} & \multirow{2}{6em}{Yes} & \SSS{} & \textbf{2.267} & 0.079 & 128.987 \\ 
& & & \MS{} & 2.416 & \textbf{0.070} & \textbf{117.278} \\ \hline
\multirow{2}{6em}{GWA \& GRIR} & \multirow{2}{6em}{78 + 94} &  \multirow{2}{6em}{No} & \SSS{} & 1.45 & 0.041 & 12.714 \\ 
& & & \MS{} & \textbf{1.135} & \textbf{0.023} & \textbf{8.412} \\ 
\end{tabular}
\caption{Results on different datasets. For BUT, GWA and GRIR datasets, test data are generated by performing convolution between anechoic speech signal and set of RIRs (1 to 6 RIRs), while for ACE dataset, babble noises (multiple people simultaneously talking) actually recorded from each room are used. For each room, 10 recordings were created for each number of sources, yielding <(\# of rooms x 10 recordings) x 6 sets> of total test recordings.} 
  \label{table:result}
\end{table*}

We present several evaluation methods to show how effective the \MS{} approach is in handling multiple source environment compared to the \SSS{} approach. Note that the results are evaluated on monaural RIRs, not binaural RIRs. 
Commonly used metrics in evaluating RIR estimation are STFT loss, T60 error and DRR error. STFT loss is the same multi-resolution loss function used during training \eqref{eq1}; T60 error is the mean squared error (MSE) of T60 values and DRR error is the MSE of DRR values between the predicted and the ground-truth RIR.
Our test data comprises of both real and simulated RIR data. For real RIR data, we used BUT \cite{szoke2019building} and ACE \cite{eaton2016estimation} datasets. From BUT dataset, we selected 3 out of 9 rooms ("Hotel\_SkalskyDvur\_Room112", "VUT\_FIT\_L227", "VUT\_FIT\_Q301") for testing. From the ACE dataset, we used the provided "babble noise" (recording of multiple people talking) as the input signal, instead of generating reverberated signals from RIRs. For simulated RIR data, we used rooms from GWA and GRIR dataset that were not used during training stage. RIRs were randomly chosen from each room to construct all test recordings. 
In order to assess both single and multi-source environments, we prepared 6 sets of test data for each dataset, starting with just one source and gradually increasing the number of sources in the recording. The final set comprises of reverberated recordings with all 6 sources playing simultaneously. 
Table \ref{table:result} summarizes the performance of the \SSS{} and \MS{} methods. The values are averages over 1 to 6 sources. We observe that training in the \MS{} setting generally improves the performance of the model on both real and simulated datasets. 


\begin{figure*}[h]
  \includegraphics[width=\textwidth]{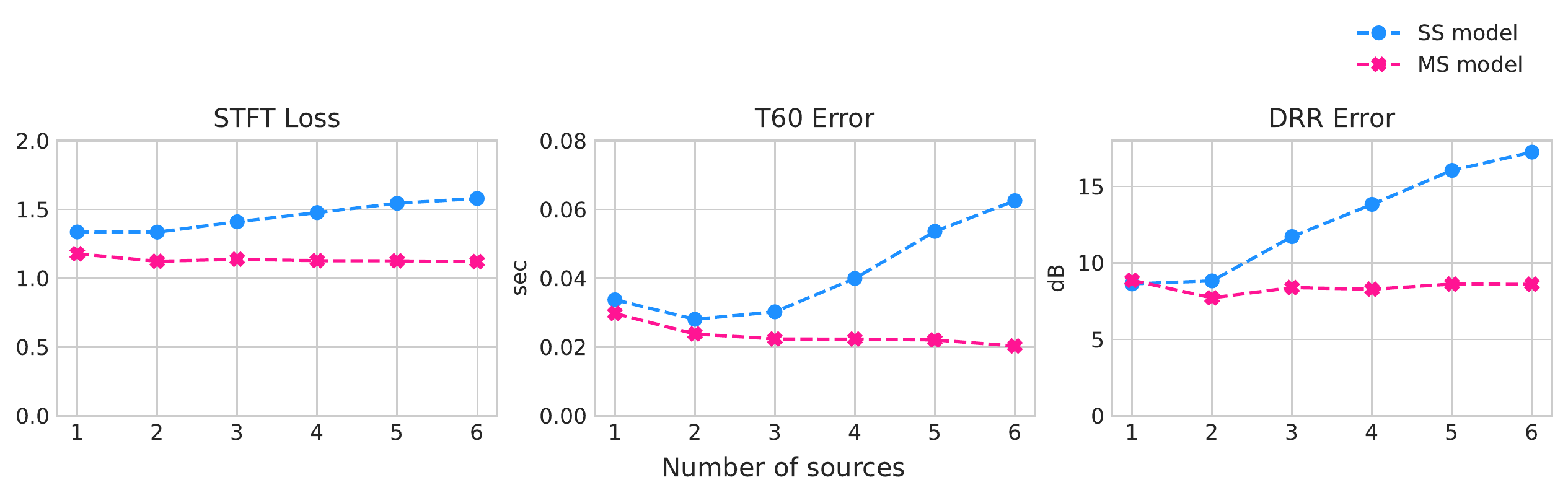}
  \caption{The performance trend of the \SSS{} and \MS{} method as the number of sources in the environment increases.}
  \label{fig:multiple_source_result}
\end{figure*}

For practical applications, it is important for a model to exhibit a stable performance, regardless of the number of sources in the environment. We plotted the performance of the model per number of sources in the input signals (Figure \ref{fig:multiple_source_result}). The evaluation shows that the model trained in the \SSS{} setting exhibits a degrading trend of performance as the number of sources increases. On the other hand, the model trained in the \MS{} setting shows a consistent performance. Importantly, we noticed that the \MS{} method performs better than the \SSS{} even in the single source environment, which implies that the \MS{} training setup can replace the \SSS{} setup for the proposed task. 

\begin{figure*}[h]
  \includegraphics[width=\textwidth]{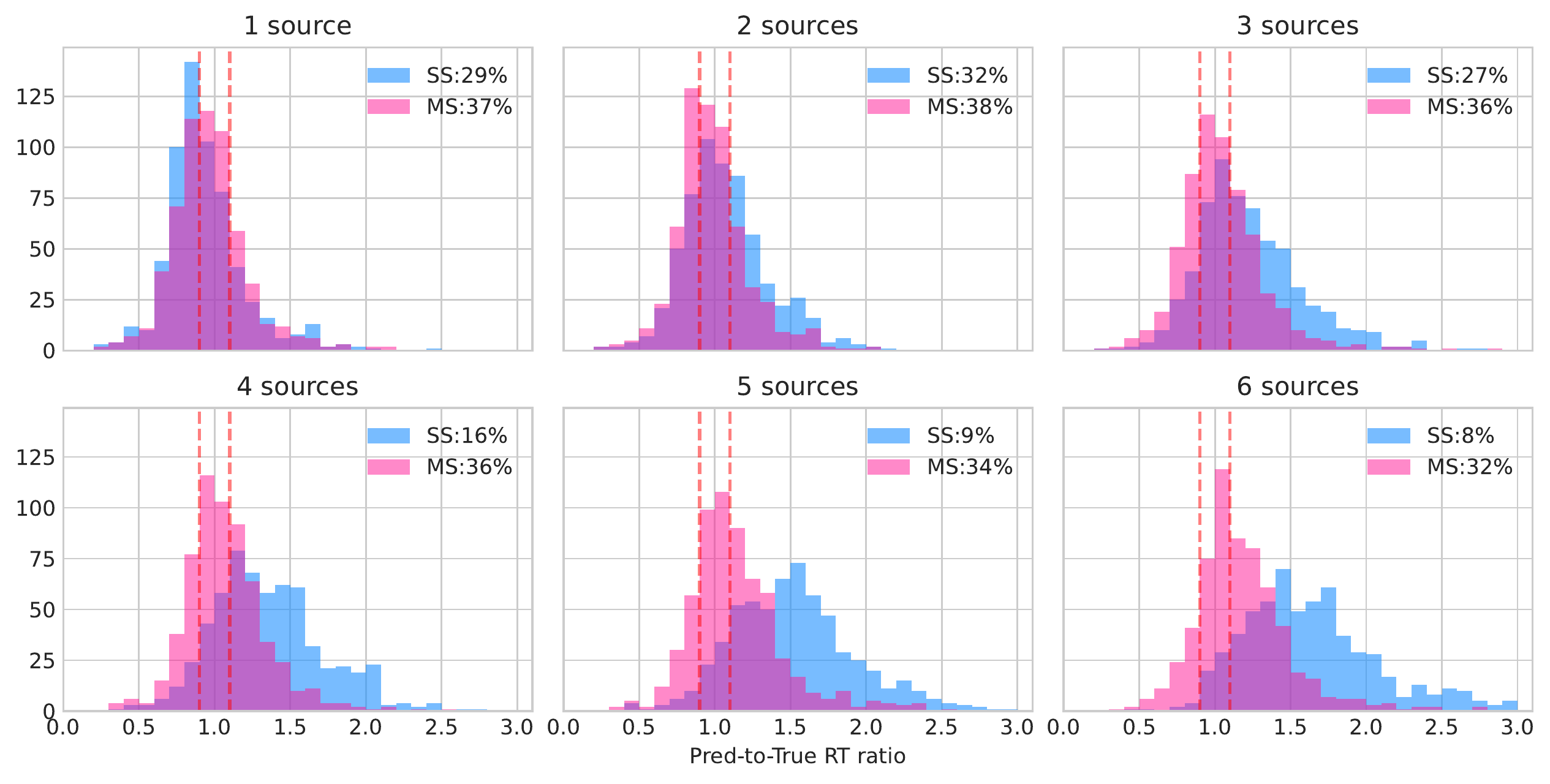}
  \caption{Distribution of pred-to-True T60 ratio for rooms with \textbf{small T60} (T60 < 0.7). Ratio of 1.0 indicates exactly correct prediction of T60. The red dashed lines represent \textpm 10\% error range and numbers in the legend box shows the percentage of rooms predicted within this perceptual range. Total of 615 recordings from GWA and GRIR test data were used.}  
  \label{fig:small_rt_ratio}
\end{figure*}

\begin{figure*}[h]
  \includegraphics[width=\textwidth]{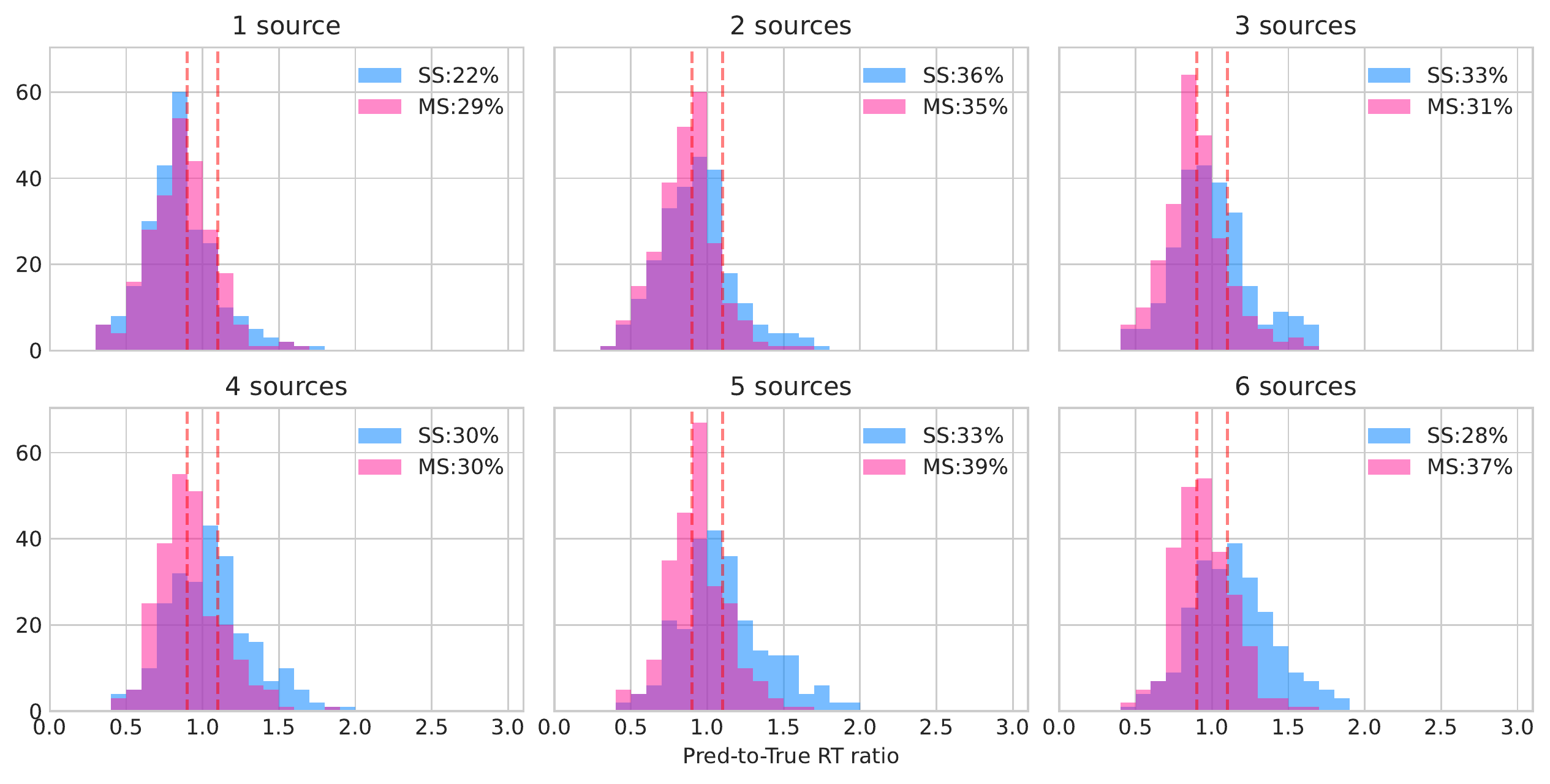}
  \caption{Distribution of pred-to-True T60 ratio for rooms with \textbf{large T60} (T60 >= 0.7). Refer to Figure \ref{fig:small_rt_ratio} for detail. Total of 245 recordings from GWA and GRIR test data were used.}
  \label{fig:large_rt_ratio}
\end{figure*}

From our observation, the \SSS{} method tends to predict larger T60 values in multiple source environments. We believe this occurs, because blind RIR estimation models most likely rely on the decay patterns to predict RIRs \cite{wen2008blind}. When there is a mixture of multiple reverberated sources, it may be more difficult to identify these patterns for individual sources. To verify this observation, we plotted the distribution of prediction-to-true T60 ratios per number of sources (Figure \ref{fig:small_rt_ratio}, \ref{fig:large_rt_ratio}). The value of 1 for prediction-to-true T60 ratio indicates that the model made an accurately prediction, while a value less than 1 indicates the model predicted the RIR with a smaller T60 value, and a value greater than 1 indicates the model predicted the RIR with a larger T60 value. Similar to Figure \ref{fig:multiple_source_result}, 
 we expect a robust model to exhibit a consistent distribution despite the number of sources in the input signal.

The prediction-to-true T60 ratios for small and large T60 rooms are shown in figures \ref{fig:small_rt_ratio} and \ref{fig:large_rt_ratio}, respectively. 
The red dashed lines represent \textpm 10\% error range, which we consider as a range of T60 values that are indistinguishable perceptually. Numbers in the legend box shows the percentage of rooms predicted within this perceptual RIR range.
Based on both figures, the \SSS{} and \MS{} methods exhibit similar distributions in the pred-to-true T60 ratio in single source scenario. However, as the number of sources increases, the \SSS{} method shows a significant shift in the direction of higher T60 ratios, especially in rooms with smaller T60 values (Figure \ref{fig:small_rt_ratio}). This supports the observation that the model tends to overestimate T60 values as the number of sources increases, leading to a sharp decrease in the percentage of correctly predicted rooms, from 27\% to 8\%. On the other hand, the \MS{} method maintains a relatively stable distribution, indicating its robustness against the number of sources. 
In highly reverberant rooms with larger T60 values, both models show no clear trend with increasing number of sources (Figure \ref{fig:large_rt_ratio}). The number of sources appear to not impact the performance of the models as much in highly reverberated environments. We assume that since identifying the decay pattern is already non-trivial in very reverberant rooms, adding more sources does not lead to more confusion. Nevertheless, the \MS{} method slightly outperforms the \SSS{} in large T60 rooms as well. 

We chose to present our results separately for rooms with small and large T60 values due to our observation that errors in T60 values are more discernible to the human ear in smaller T60 rooms. Therefore, showing the model's robustness in such rooms would be more valuable. By plotting the rooms separately, we were able to observe the distinct trend of performance in small and large T60 rooms and demonstrate that the \MS{} model exhibits more promising performance in small T60 rooms.

\subsection{Qualitative evaluation}
We internally carried out a MUSHRA listening test with 16 participants. Participants were asked to rate the similarity between the reference signal, which was a ground truth RIR convolved with an anechoic signal, and the testing signals, which were predicted RIRs from the \SSS{} and \MS{} methods convolved with anechoic signals. The test consisted of 3 questions per room, and  there were a total of 3 rooms. For each room, 2 questions were simulating a multiple source scenario and 1 question was simulating a single source scenario. The rooms selected had an average T60 value of 0.37, 0.85, and 1.31. 

\begin{figure}[h]
  \includegraphics[width=\columnwidth]{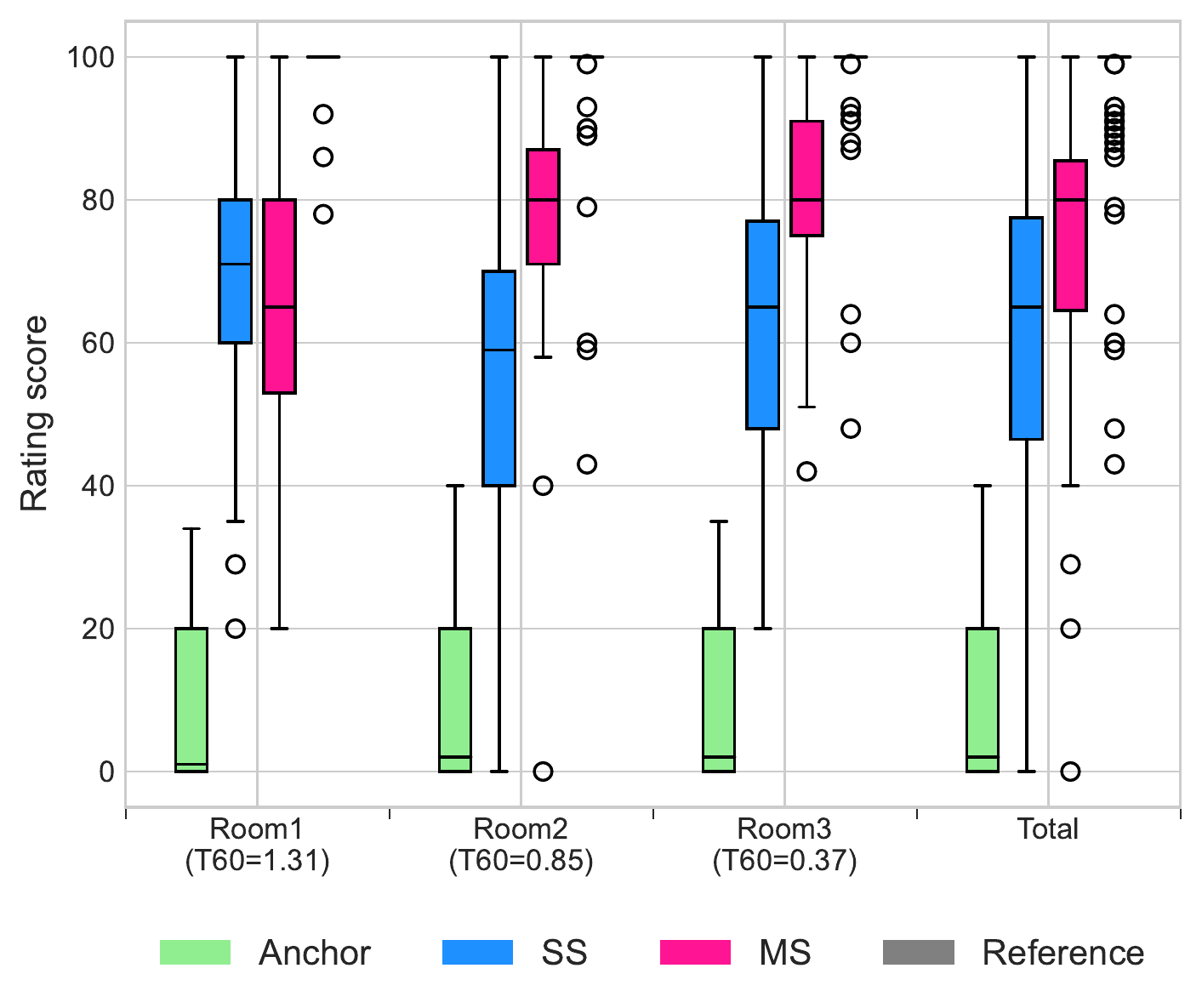}
  \caption{MUSHRA test result}
  \label{fig:listening_test}
\end{figure}

Figure \ref{fig:listening_test} shows the results for each room, as well as for all three rooms ("Total"). While for Room 1, the \MS{} method appears to be rated slightly lower than the \SSS{} method, for other 2 rooms, the \MS{} method is given visually higher scores. To formalize the result with the statistical significance, we performed the Kruskal-Wallis H-test, which confirmed that there are significant differences among the median values of the ratings (F=413.13, p << 0.001). Additionally, we performed a pairwise post-hoc pairwise test with Conover's test, which also showed that there is a significant difference between the \MS{} and the \SSS{} method (p << 0.001). For the first room, we performed separate post-hoc pairwise test and obtained a result that there is no statistically significant difference between the \MS{} and the \SSS{} method (p=0.080). 
In conclusion, the listening test verified that users generally perceived the predicted RIRs from the \MS{} method to be closer to the ground truth RIR.

\section{Summary}
In this work, we presented a training method for blind RIR estimation that is able to handle signals in a multiple source environment. We showed that by training the model to estimate the representative RIR, the model is able to predict stable RIR regardless of the number of sources in the input signal. Thus, the proposed method is more robust under multiple source environments, compared to the previous single source training methods. 

While the current method provides a promising direction, there still is much room for improvement and exploration of the model and the training method. One of the major limitation of the model used in this work is that it assumes the boundary between the early reflection and the late reverberation to be around 0.05s. Designing a model that handles varying boundary location, depending on the T60 values, can be one possible option. While the task introduced in this study was to estimate a single representative RIR, another direction is to identify the number of sources and estimate RIRs for each of the sources, which will be especially useful for modeling individual sources in virtual reality applications.

\bibliographystyle{jaes}

\bibliography{refs}

\end{document}